\documentclass[referee]{aa} 

\usepackage{epsfig,lscape} 
\usepackage[dvipdfm]{hyperref} 
\usepackage{natbib}
\usepackage{graphicx}
\usepackage[varg]{txfonts}

\begin{document}

\titlerunning{Characterizing AGB stars in WISE bands}
\authorrunning{J.H. Lian et al.}

\title{Characterizing AGB stars in Wide-field Infrared Survey Explorer (WISE) bands \footnote{The synthetic photometry and input papameters for our model grid are only available in elctronic form at the CDS via anonymous ftp to cdsarc.u-strasbg.fr (130.79.128.5) or via http://cdsweb.u-strasbg.fr/cgi-bin/qcat?J/A+A/}}
\author{Jianhui Lian
        \inst{1,2} 
\and Qingfeng Zhu
     \inst{1,2}
\and Xu Kong
     \inst{1,2} 
\and Jinhua He\inst{3}}

\institute{Center for Astrophysics, University of Science and Technology
of China, Hefei 230026, China\\
\email{ljhhw@mail.ustc.edu.cn; zhuqf@ustc.edu.cn; xkong@ustc.edu.cn} 
\and Key Laboratory for Research in Galaxies and Cosmology, Chinese Academy of Sciences, Hefei 230026, China 
\and Key Laboratory for the Structure and Evolution of Celestial 
Objects, Yunnan Observatories,
CAS, Kunming 650011, China}

\abstract
{}
{Since asymptotic giant branch (AGB) stars are bright and extended infrared objects, most Galactic AGB stars saturate the
Wide-field Infrared Survey Explorer (WISE) detectors and therefore the WISE magnitudes that are restored by applying point-spread-function fitting need to be verified.
Statistical properties of circumstellar envelopes around AGB stars are discussed on the basis of a WISE AGB catalog verified in this way.}
{We cross-matched an AGB star sample with the WISE All-Sky Source Catalog and the Two Mircon All Sky Survey catalog. Infrared Space Observatory (ISO) spectra of a subsample of WISE AGB stars were also exploited.
The dust radiation transfer code DUSTY was used to help predict the magnitudes in the W1 and W2 bands, the two WISE bands most affected by saturation, for calibration purpose, and to provide physical parameters of the AGB sample stars for analysis.}
{DUSTY is verified against the ISO spectra to be a good tool to reproduce the spectral energy distributions of these AGB stars.
Systematic magnitude-dependent offsets have been identified in WISE W1 and W2 magnitudes of the saturated AGB stars,
and empirical calibration formulas are obtained for them on the basis of 1877 (W1) and 1558 (W2) AGB stars that are successfully fit with DUSTY. 
According to the calibration formulae, the corrections for W1 at 5 mag and W2 at 4 mag are $-0.383$ and 0.217 mag, respectively. 
In total, we calibrated the W1/W2 magnitudes of 2390/2021 AGB stars.
The model parameters from the DUSTY and the calibrated WISE W1 and W2 magnitudes
are used to discuss the behavior of the WISE color-color diagrams of AGB stars. The model parameters also reveal that O-rich AGB stars with opaque 
circumstellar envelopes are much rarer than opaque C-rich AGB stars toward the anti-Galactic center direction, 
which we attribute to the metallicity gradient of our Galaxy.}
{}


\maketitle

\section{introduction}
The asymptotic giant branch (AGB) phase is the final stellar evolutionary
stage of intermediate-mass (1 -- 8 $M_{\odot}$) stars driven by nuclear
burning. Stars in this stage have low surface effective temperatures
(below 3000 K) and experience intense mass loss (from $10^{-7}$ to
$10^{-4}$ ${M}_{\odot}{\rm yr}^{-1}$) \citep{Habing1996}. Heavy elements in the
mass outflow from a central star will condense to form dust when the gas
temperature drops to the sublimation temperature of the dust
grains. Dusty circumstellar envelopes will form at the distance of
several stellar radii. Dust grains in the envelopes absorb stellar
radiation and re-emit in the infrared. Thus, AGB stars are important
infrared sources. The mass-loss process plays an important role in the
evolution of AGB stars because it affects the lifetime of the AGB
phase and the core-mass of the subsequent post-AGB stars. Statistics
of a large sample of AGB stars would help to constrain the evolution
of dust envelope. Due to the relative over-abundance between carbon
and oxygen, there are two main types of AGB stars: (1) the O-rich with
C/O $<1$ and mainly silicate-type grains in the outflow, and (2)
C-rich with C/O $>1$ and mainly carbonaceous grains in the
envelopes. The different dust compositions of these two types of
  AGB stars result in different infrared spectral features, which can be used to distinguish the two groups of the stellar objects.

The first-generation infrared space-telescope, the Infrared Astronomical
Satellite ({\sl IRAS}; \citealt{Neugebauer1984}), mapped the sky in two
mid-infrared and two far-infrared bands (12, 25 $\mu$m and 60, 100
$\mu$m, respectively) in the 1980s. It discovered many mass-losing AGB
stars in the Milky Way and the Magellanic Clouds. Several dozen AGB
stars were observed spectroscopically by the Infrared Space Observatory
\citep[ISO;][]{Kessler1996} with the Short-Wave Spectrometer (SWS) onboard, which works in the range of 2.4 $\mu$m to 45 $\mu$m. These infrared spectra
provide tremendous information about dusty circumstellar envelopes around
the AGB stars, but they are limited by the sample size. The Galactic Plane Survey of the Spitzer Space Telescope \citep{werner2004} provides high-resolution and sensitive infrared images of Galactic plane at the 3.6, 4.5, 5.8, 8.0, 24, and 70 $\mu$m bands. However, it cannot adequately separate the two types of AGB stars in mid-infrared color-color diagrams without filters between 8 $\mu$m and 24 $\mu$m. Launched 
on 2009 December 14, the Wide-field Infrared Survey Explorer (WISE)
completed an entire sky survey in the 3.4, 4.6, 12 and 22 $\mu$m bands
(hereafter named W1, W2, W3, and W4, respectively) in 2010. The
FWHMs of the point spread functions (PSF) are $6.''1$, $6.''4$, $6.''5$, and
$12.''0$ for the four WISE bands, and the sensitivities ($5 \sigma$) of the point sources are 0.08, 0.11, 1, and 6 mJy \citep{Wright2010}. The Two Micron All Sky Survey (2MASS; \citealt{Skrutskie2006}) also mapped the entire sky in three near-infrared, J (1.25 $\mu$m), H (1.65 $\mu$m) and $\mathrm{K}_{\rm s}$ (2.17 $\mu$m) bands. They are an important supplement to the mid-infrared WISE data to complete the infrared spectral
energy distribution (SED) of AGB stars.

Recently, galaxy-wide AGB populations have been identified and studied in nearby galaxies using ground- and space-based near- and mid-infrared data (\citealt{riebel2012}, \citealt{boyer2012} in the Magellanic Clouds and \citealt{javadi2013} in M33). While similar research in our galaxy is hindered by foreground extinction, there are some studies on large datasets of Galactic AGB stars, such as the work of \citet{jura1989} in the solar neighborhood. 
\citet{Suh2009,Suh2011} composed a large verified AGB catalog in our galaxy from the literature and investigated infrared colors of AGB stars and their distribution in color-color diagrams. To study an AGB star population, a more time-saving way is constructing a set of model grids to simulate the infrared SED of AGB stars and acquire their physical parameters, such as mass-loss rate and optical depth. Many model grids have been developed to study the evolved-star populations in the local group galaxies \citep{groenewegen2006,sargent2011,srinivasan2011}. All of the models are publicly available. However, the model set of \citet{groenewegen2006} does not have WISE synthetic photometry. The Magellanic Clouds have a lower metallicity than our Galaxy, and \citet{sargent2011} and \citet{srinivasan2011} chose a stellar photosphere model with subsolar metallicity as input of a radiative transfer model, which is not the case in our galaxy. Therefore, we used DUSTY \citep{ivezic1997} to generate a new grid of radiative transfer models to simulate the infrared SEDs of Galactic AGB stars.

As AGB stars typically are bright IR objects, most of our AGB samples are brighter than the WISE detector saturation limits and only photometry of unsaturated pixels are available, therefore their WISE magnitudes need to be verified. The main goal of this paper is to study the properties of AGB stars in WISE bands and recalibrate the WISE photometry for sources saturated in WISE bands. We describe our AGB sample, DUSTY
algorithm and calibration method in \textsection2. A possible
calibrating solution is presented in \textsection3.1. We present
our results about the distinct location of the two types of AGB stars in a color-color diagram in \textsection3.2, the model parameter effects on WISE colors in \textsection3.3, and the Galactic longitude distribution in \textsection3.4. Our summary is presented in \textsection4. All the magnitudes mentioned in this paper are Vega magnitudes.

\section{Sample and method}
\subsection{AGB sample}

\citet{Suh2009} composed a catalog of Galactic AGB stars based on reports of the {\sl IRAS} Low Resolution Spectrograph (LRS; $\lambda=8$-22 $\mu$m), ISO Short-Wave Spectrometer (SWS), the Near Infrared Spectrometer (NIRS; 1.4-4.0 $\mu$m; \citealt{murakami1994}), radio OH and SiO survey \citep{lewis1990}, the Midcourse Space Experiment (MSX; \citealt{egan2003}) and the 2MASS data. \citet{Suh2011} (hereafter Suh11) updated the catalog with SiO maser sources for O-rich AGB stars and additional sources for C-rich AGB stars. The updated catalog contains 3003 O-rich and 1168 C-rich AGB stars with their IRAS colors and has the largest confirmed AGB collection so far. This catalog provides an important entry for the future study of AGB stars.

To study the properties of AGB stars in infrared bands, we
cross-correlated the AGB catalog with the WISE All Sky Catalog. Because
{\sl IRAS} data have a positional uncertainty ellipse of
$45^{\prime\prime}$$\times{9^{\prime\prime}}$, which is much bigger
than the position uncertainty in WISE photometry ($\sim$
$0.''20$), we used coordinates and magnitudes of
sources during the cross-correlation to improve the precision. We
first used $45''$ as the searching radius to find matching
candidates. In the cases of multiple matches, we selected the WISE
source with the highest 12 $\mu$m flux. AGB stars are among the
brightest sources in the sky at mid-infrared wavelengths. The
detection limit of {\sl IRAS} is higher than that of WISE. All 12 $\mu$m
{\sl IRAS} sources needed also to be detected by WISE. However, high flux
saturates the WISE detectors and may render them unreliable. A
selection criterion of W3 fainter than $-2$ mag was applied to exclude
other 70 O-rich and 33 C-rich AGB stars from the sample. This
criterion is based on our analysis of WISE magnitudes and ISO SWS
spectra of AGB stars, which is discussed in
\textsection 2.3. The 2MASS observations
in J, H and $\mathrm{K}_{\rm s}$ bands are included in the WISE catalog. For each WISE
source, the associated 2MASS observations were selected with the
nearest sources in the 2MASS catalog within $3''$. 
Some of these AGB stars (730 O-rich and 158 C-rich) do not have a 2MASS counterpart, which is perhaps due to heavy circumstellar extinction. This tentative interpretation is supported by the fact that their W3--W4 colors (with medians of 1.874 mag for O-rich and 0.993 mag for C-rich objects), which correspond to an optical depth $\sim$ 40 at 0.55 $\mu$m (according to our discussion in \textsection3.2), are much redder than that of the whole AGB sample (W3--W4=1.314 mag for O-rich and 0.454 mag for C-rich objects).
Finally, 2203 (73.6\%) O-rich and 958 (83.6\%) C-rich AGB
stars with 2MASS observations remained. They are our WISE AGB sample. 
We did not apply interstellar extinction correction to our data because there are no distance estimates. The extinction caused by the interstellar material (ISM) can be included in the extinction of the dust envelope around AGB stars. We expect that the optical depth is overestimated by a factor depending on the distance to the Sun and specific sightlines. 
The good agreement between DUSTY models and ISO data suggests that our results from estimating the saturated WISE W1 and W2 photometry are not affected by interstellar extinction significantly.

To calibrate the WISE photometry, we 
cross-matched the Suh11 catalog with sources observed by ISO SWS
(spectrum data reduced by \citealt{sloan2003}). To obtain a reliable
spectral sample, we excluded several O-rich AGB star spectra that
deviate significantly from the typical SEDs of AGB stars and
lack 10 $\mu$m silicate feature. Spectra with a low signal-to-noise ratio were also excluded. Finally, we obtained 67 O-rich and 44 C-rich AGB stars with reliable ISO mid-infrared spectra. Among them, 66 (49) O-rich and 44 (37) C-rich AGB stars have associated WISE (and 2MASS) observations.

\subsection{Theoretical modeling}

DUSTY is a radiative transfer code developed by 
  \citet{ivezic1997} and can be used to model the dusty circumstellar
envelope around AGB stars. Many authors have used this code to simulate
the infrared emission of AGB stars \citep[e.g.,][]{Riechers2004,
  Matsuura2006, Groenewegen2012}. Taking the advantage of scaling,
\citet{ivezic1997} minimized the number of input parameters by
assuming spherical symmetry. The parameters that have to be specified
in models are (1) the input radiation field, which can be described by
the effective temperature $T_{\scriptsize{\textup{eff}}}$ of the cental star; (2) the optical properties of dust grains with a specified
chemical composition; (3) the grain-size distribution $n(a)$; (4) the
dust temperature at the inner boundary of the circumstellar envelope
$T_{\scriptsize{\textup{in}}}$; (5) the relative thickness (radius at
the outer boundary over radius at the inner boundary,
$r_{\scriptsize{\textup{out}}}/r_{\scriptsize{\textup{in}}}$) of the
envelope; (6) envelope density distribution assuming a power law
of $\rho(r) \propto r^{-\alpha}$ in this work; (7) the overall optical depth at a reference wavelength $\tau_{\lambda}$, and we assumed a reference wavelength at 0.55 $\mu$m.

We used DUSTY to generate model templates to simulate the infrared SEDs of AGB stars in our Galaxy. For the radiative field, we took the COMARCS hydrostatic models \citep{aringer2009} of AGB star photospheres to represent the C-rich AGB stars and PHOENIX models \citep{kucinskas2005} for O-rich AGB stars. Both models assume metallicity $Z=Z_{\odot}$. 
As noted in \citet{aringer2009}, most C-rich AGB stars have temperatures lower than 3500 K; the effective temperature of C-rich stellar models in this work ranges from 2500 K to 3500 K with increments of 500 K. For O-rich AGB stars, we used PHOENIX models with stellar effective temperatures between 2100 K and 4500 K in increments of 400 K. The effective temperature increments are larger than in the model grids of \citet{sargent2011} and \citet{srinivasan2011}, which are 200 and 100 K, respectively, to save calculation time. We explored the effective temperature with increments as small as 100 K and did not find a noticeable increase in the number of AGB stars with good fits. We set the stellar mass for O-rich AGB stars to be 1 $M_{\odot}$, which is the only available choice in the PHOENIX model and a better choice than a blackbody SED. As shown in \citet{aringer2009} and \citet{kucinskas2005}, the surface gravities (log(g/(cm ${\rm s}^{-2}$))) and C/O ratio have a minor effect on near-infrared colors, therefore we only used COMARCS models with log(g)=--0.4 (when the temperature was higher than 3200 K, only log(g)=--0.2 is available) and C/O=1.1, and PHOENIX models with log(g)=--0.5 and C/O=1.1.

Suh11 also used DUSTY to simulate the infrared
colors of AGB stars. They assumed that each type of AGB circumstellar
envelope contains only two types of dust grains. The circumstellar
envelopes of O-rich AGB stars only contain silicate and porous
corundum $\textup{Al}_{2}\textup{O}_3$. The circumstellar envelopes of
C-rich AGB stars contain amorphous carbon (AmC) and SiC. A composition
ratio parameter X was defined to specify the number ratio of the two types
of dust particles. In O-rich AGB stars,
$\textup{X}=\textup{silicate}/(\textup{silicate}+\textup{Al}_{2}\textup{O}_3)$;
in C-rich AGB stars,
$\textup{X}=\textup{AmC}/(\textup{AmC}+\textup{SiC})$. They
found that this assumption can well describe the properties of the
envelopes of AGB stars in general. In many previous studies of radiative transfer models of AGB stars, the dust composition ratio X is fixed. For example, \citet{sargent2011} used 100\% silicate for O-rich AGB stars and \citet{srinivasan2011} used a mixture consisting of 90\% AmC and 10\% SiC for C-rich AGB stars. \citet{groenewegen2006} introduced three dust composition ratios (X=100\%, 40\% and 0\%) for O-rich and two ratios (X=100\% and 85\%) for C-rich envelopes. In this work, to save time and for completeness, we fixed the dust ratio to 100\% silicate around O-rich AGB stars and adopted five dust composition ratios of AmC and SiC ranging from 100\% to 60\% for C-rich stars. The optical properties of silicate grains were taken from \citet{ossenkopf1992} 
and the porous corundum
$\textup{Al}_{2}\textup{O}_3$ from \citet{Begemann1997}. While the optical properties of porous corundum from \citet{Begemann1997} only cover a wavelength range redward of 7.8 $\mu$m, the refraction indices are assumed to be constant at shorter wavelengths, with a value equal to the corresponding end point. We can see from Figure 2 of \citet{Begemann1997} that the refractive index from \citet{koike1995} and \citet{chu1988} at wavelengths shorter than 7.8 $\mu$m is roughly constant. 
The optical properties of AmC were taken from \citet{Suh2000} and SiC from
\citet{Pegourie1988}. The grain-size distribution was modeled as $n(a) \propto a^{-q}$ for $a_{min} < a < a_{max}$ (MRN distribution, \citealt{Mathis1977}) with a power-law index q=3.5, $a_{min}=0.005\,\mu$m and $a_{max}=0.25\,\mu$m.

In DUSTY models, $T_{in}$ is specified directly instead of the inner radius. We calculated models with $T_{in}$ ranging from 600--1800 K for C-rich and 600--1400 K for O-rich AGB stars with increments of 200 K. Dust grains cannot form at temperatures higher than the sublimation temperature, which is approximately 1400 K (1800 K) for silicate (graphite grains) \citep{posch2007,speck2009}. The ranges of $T_{in}$ are roughly consistent with that of models in \citet{sargent2011} and \citet{srinivasan2011}. We assumed that the power of the density profile $\alpha=2$ for
the circumstellar envelopes, which is typical for steady
  circumstellar winds. When $\alpha=2$, we find that the variance of the relative thickness of the shell from $10^2$ to $10^5$ does not have a significant impact on the mid-infrared colors at WISE wavelengths. Thus, the relative thickness was fixed at $4 \times 10^3$. The optical depth $\tau$ at 0.55 $\mu$m ranges from $10^{-2}$ to $10^2$ and is sampled in logarithmic space, with a finer stepsize at the highest optical depths to ensure an even distribution of models in color-color space. Finally, we calculated 3 (7) values of $T_{\textup{eff}}$, 5 (1) values of X, 7 (5) values of $T_{\textup{in}}$, and 29 (29) values of $\tau_{0.55}$ for C-rich (O-rich) AGB stars. This gives a total of 3045 and 1015 models for C-rich and O-rich AGB stars, respectively. The ranges of model parameters are listed in Table 1.

Given each set of parameters of $T_{\scriptsize{\textup{eff}}}$, X, $T_{\scriptsize{\textup{in}}}$, $r_{\scriptsize{\textup{out}}}/r_{\scriptsize{\textup{in}}}$, $\alpha$, $\tau_{\scriptsize{\lambda}}$, DUSTY computes the emerging SED (from 0.1 $\mu$m to 3600 $\mu$m) from the envelope. The SED consists of three components: attenuated stellar radiation, scattered radiation, and dust emission. The emerging SED can be convolved with WISE photon-counting relative system response (RSR) curves, $\lambda R(\lambda)$ \citep{jarrett2011}, to simulate the WISE photometry:
\begin{equation}
W_n=-2.5 \mbox{log} \left(\frac{\int F_{\lambda} \lambda R_{\lambda} d\lambda}{F_{\lambda n} b_n} \right)  \qquad n=1,2,3,4,
\end{equation}
Here $b_n$ is the width (in units of $\mu$m) of each WISE band. $F_{\lambda n}$
are the fluxes at the magnitude zero point in the four bands. Since WISE saturates on Vega, the magnitude zero points are based on fluxes of fainter stars calibrated to the Vega system. We applied 2MASS relative spectral response curves derived by \citet{Cohen2003} to obtain the simulated 2MASS magnitudes. This procedure was used to simulate the WISE and 2MASS magnitudes for each model template.

\begin{table}
\caption{Parameter range for model templates}
\label{table:1}
\centering
\begin{tabular}{l c c c c}
\hline\hline
             &  O-rich range & increment  & C-rich range & increment \\
\hline
$T_{eff}$ (K)    & 2100 --- 4500  & 400    &  2500 --- 3500  & 500  \\
X            & 100\%    &  ... & 100\% --- 60\% & 20\% \\
$T_{in}$ (K)     & 600 --- 1400  & 200   &  600  --- 1800   &   200  \\
log($\tau_{0.55}$)   & $-2$ --- 2 &  0.3$^a$    &  $-2$ --- 2  & 0.3$^a$   \\
\hline
\end{tabular}\\
Notes:
$^a$ The increments for log($\tau_{0.55}$) in intervals of [$-$2, 0.7],[0.8, 1.4], and [1.45, 2] are 0.3, 0.1, and 0.05, respectively.
\end{table}

\subsection{WISE photometry and calibration method}

We applied the WISE RSRs to the ISO spectra to
simulate WISE photometry for these ISO objects. Figure 1 compares
  the observed WISE magnitudes with those simulated from the ISO
spectra. In the figure, the 66 O-rich stars are designated as
plus signs and the 44 C-rich AGB stars as triangles. Dashed
lines indicate equality between the x- and y-axis values. Solid lines in W3 and W4 panels are the best-fit relations given by 
\begin{eqnarray}
W3(ISO) & = & -0.13(\pm 0.26)  \nonumber \\
& & +1.00(\pm 0.00)\times W3(WISE) \nonumber \\
W4(ISO) & = & -0.15(\pm 0.09) \nonumber \\
& & +0.95(\pm 0.09)\times W4(WISE). 
\end{eqnarray}
We fit the line equation for W3 with a fixed unity slope to safely extrapolate the relation to fainter magnitude. 
The correlation between ISO
and WISE is quite good in the W3 and W4 bands, except that a turning
point exists in the W3 band. The correlation only holds for objects
fainter than $-2$ mag in W3. That is the reason why we excluded sources
with W3 brighter than --2 when we assembled our WISE AGB sample in
\textsection 2.1. As ISO and WISE observations were taken at different epochs and AGB stars are variable objects even in infrared bands, part of the spread in the W3 and W4 equation is due to the variability of AGB stars. 
We also compared WISE W3 and W4 bands with IRAS 12 $\mu$m and 25 $\mu$m data and found they agree well. However, no obvious correlation is observed between WISE data and ISO observations in the W1 and W2 bands.

\begin{figure}
\resizebox{\hsize}{!}{\includegraphics{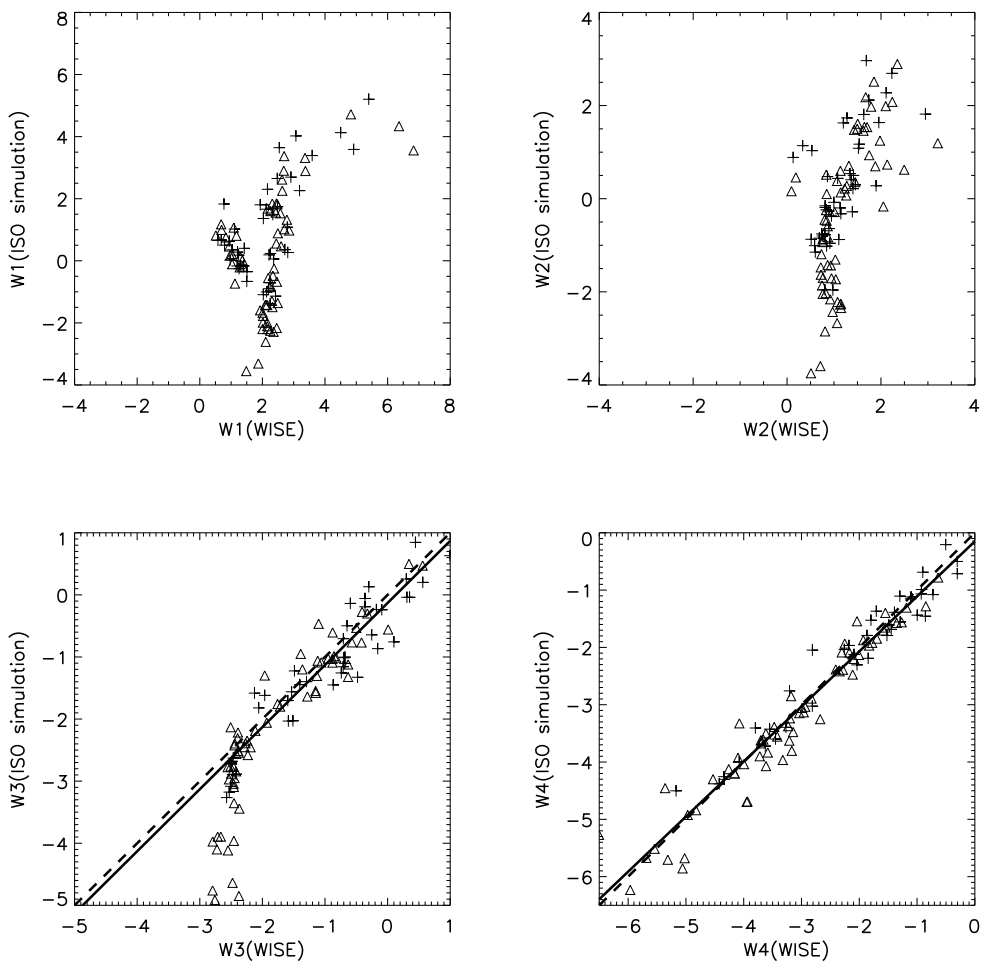}}
\caption{Comparison between the WISE magnitudes and the ISO spectral observations of AGB stars in WISE W1--W4 bands. Pluses are O-rich and triangles are C-rich AGB stars. Dashed lines indicate equality between the x- and y-axis values, and solid lines are the best-fit linear relations where applicable.}
\label{figure1}
\end{figure}

It is mentioned in the WISE Explanatory Supplement documentation of the
All-Sky Data Release that bright sources will saturate WISE detectors
\footnote{WISE Explanatory Supplement documentation for All-Sky Data
  Release
  http://wise2.ipac.caltech.edu/docs/release/allsky/expsup/}. The
saturation limits are 8.1, 6.7, 3.8, and --0.4 magnitudes for the W1-W4 bands,
respectively. For saturated objects, WISE fits the
point spread function (PSF) to the unsaturated pixels on the images
to recover the saturatured pixels and yield photometry for these objects (we call this PSF-fit
photometry). The bright source photometry limits for the WISE 4 bands are 2, 1.5, --3, and --4 mag. When they are brighter than this limit, all the pixels are saturatured and no unsaturated pixels are available to fit to the PSF. Therefore the PSF-fit photometry is quite unreliable for sources brighter than the photometry limit. Because AGB stars typically are bright IR objects, most of
our AGB samples are brighter than the saturation limits, and therefore only
PSF-fit photometry is available. This comparison suggests that 
the PSF-fit photometry can be directly calibrated to the ISO synthetic photometry in the W3 and W4 bands and becomes quite unreliable for
almost all the considered bright ISO AGB stars in W1 and W2 bands. 
Our ISO samples are confined to a relatively bright synthetic W3 magnitude range of $W3<0.5$. To facilitate our calibration down to the WISE saturation limit of 3.8 mag, we assumed that the good correlation between the PSF-fit and ISO synthetic W3 photometry can be extrapolated. To ensure a reliable extrapolation, we adopted the fit with unity slope (Eq.2). To
compensate for the loss of the W1 and W2 band data for bright objects like our AGB
stars, we developed a calibration strategy based on the use of
the DUSTY model template introduced in \textsection2.2.

To find the best-fit model to each source, we adopted the following fitting procedure: the goodness of fit was measured by the overall offset between observed and model magnitudes. For each object, each set of model magnitudes was shifted together to obtain the smallest offset from the observed magnitudes. The overall shift is then the average of the differences in the five bands. We assumed that an overall shift smaller than 0.4 mag consistutes a good fit. For comparison, the average root mean square observational uncertainties at these five bands are 0.165 and 0.198 mag for O-rich and C-rich AGB, respectively.

We obtained the W3 and W4 magnitudes of the ISO AGB sources by convolving
ISO spectra with proper filter-response curves. 
Figure 2 shows the ISO spectra as a solid line and the associated 2MASS J, H, $\mathrm{K}_{\rm s}$
fluxes (squares) for one AGB star of each type. The dotted
lines are the fitted model SEDs. The models fit
the observations well in the bands of interest. The best-fit model parameters and deviation from observations are displayed in each panel. We note that the best-fit model does not fit the dust resonance features (such as 9.7 $\mu$m and 18 $\mu$m silicate features) very well, since we only have two synthetic photometry points to sample this part of the SED. There are some other
wavelength regions where model spectra do not match the observations, such as
the absorption feature at about 5 $\mu$m in the spectrum of the
selected C-rich AGB star, which is absent from the model SED. The absorption feature at about 5 $\mu$m may be the molecular features of $C_3$ at 5.1 $\mu$m and/or CO at 4.6 $\mu$m \citep{gautschy2004}. The carbon-star spectrum shown in Figure 2 also has deep features at $\sim$ 3 $\mu$m and $\sim$ 14 $\mu$m, which are caused by HCN+$\textup{C}_2\textup{H}_2$ \citep{gautschy2004}. Strong $\textup{C}_2\textup{H}_2$ features have been observed in carbon stars in nearby galaxies; in the Milky Way's high metallicity it is expected that there is a larger contribution from HCN to these features \citep{Matsuura2006}. 
More complicated DUSTY models with additional dust and molecular species may improve our fits. However, this requires significant efforts on choosing dust compositions for models and beyond the scope of the current work. The difference of the W2 magnitude of this C-rich AGB star between observation and model is 0.11 mag. According to our analysis, this level of uncertainty does not significantly affect our results and our conclusions are still valid. We expect that the model spectra and dust properties can be better constrained with more photometry points in some narrow-band filters that trace the molecular and dust features.

\begin{figure}
\resizebox{\hsize}{!}{\includegraphics{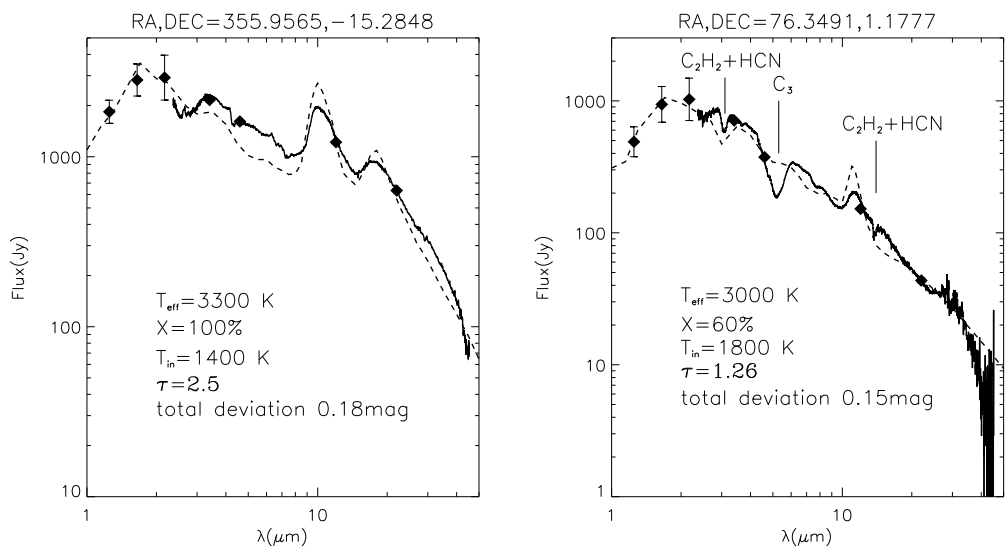}}
\caption{Comparison between photometric/spectroscopic observations and
  DUSTY best-fit models for AGB stars. The left panel is for a selected
  O-rich star and the right panel for a C-rich star. The solid lines are ISO
  SWS spectra in the range between 2.5 $\mu$m and 45 $\mu$m. The
  diamonds represent 2MASS J, H, and ${\rm K_s}$-band fluxes and ISO synthetic photometry for W1, W2, W3, and W4. The dotted lines are the best-fit models described in the text. The molecular features discussed in the text are marked.}
\label{figure2}
\end{figure}

In total, 45 O-rich AGB and 28 C-rich AGB stars with ISO spectra were
successfully modeled. The comparison of W1 and W2 magnitudes between
ISO observations and the DUSTY models are shown in Figure 3. Dashed
lines indicate equality between the x- and y-axis values. The data points are fitted with a linear equation as 
\begin{eqnarray}
W1(ISO)=-0.21(\pm 0.07)+1.09(\pm 0.05)\times W1(model) \nonumber \\
W2(ISO)=-0.15(\pm 0.06)+1.08(\pm 0.05)\times W2(model).  
\end{eqnarray}
The correlations of both O-rich and
C-rich AGB stars (represented by pluses and triangles, respectively) are very close to the one-to-one equation, which suggests that the DUSTY models agree well with the ISO observations despite the degeneracy of some model parameters and no correction for interstellar extinction. The slight deviation from ISO synthetic photomety may be caused by the molecular features, which are absent in the model spectra.

Based on the above results, we calibrated our WISE AGB sample in five
steps. First, we calibrated WISE W3 and W4 PSF-fit photometry to ISO synthetic photometry by using Equation 2. Second, we fit the calibrated   W3 and W4 and the 2MASS J, H, and $\mathrm{K}_{\rm s}$-band
magnitudes and determined the best DUSTY model for each WISE AGB
star. Third, we compared the model W1 and W2 magnitudes with the
  observed ones for our whole sample of AGB stars
  and identified the magnitude ranges in which good linear correlations
  hold. Fourth, we assumed that the ISO simulated W1 and W2
  magnitudes are reliable as a whole to derive empirical
formulas for calibrating the observed W1 and W2 magnitudes to those of the DUSTY model
and then to those simulated by ISO. Finally, we applied the
obtained calibration formulas to the observed W1 and W2 magnitudes of
the objects in the good-correlation magnitude ranges. As a result,
1659 of 2203 O-rich and 835 of 958 C-rich WISE AGB stars meet our good-fit criteria. In total, we calibrated the W1/W2 magnitudes of 2390/2021 AGB stars. The detailed calibration solution is discussed in \textsection3.1. We note that about one fourth (544) of the O-rich AGB stars cannot be fitted by our model grid. We explored the model grid from \citet{sargent2011} and found that the situation is not improved. Most of the objects without best-fits tend to have higher W3--W4 colors than fitted ones. Some even have an extreme W3--W4 color with strong deviations from the main distribution of O-rich AGB stars in the color-color diagram and seem to be misclassified as AGB stars or have some extreme properties that we do not know yet. We explored many possible model parameter ranges and did not find a significant improvement.
\begin{figure}
\resizebox{\hsize}{!}{\includegraphics{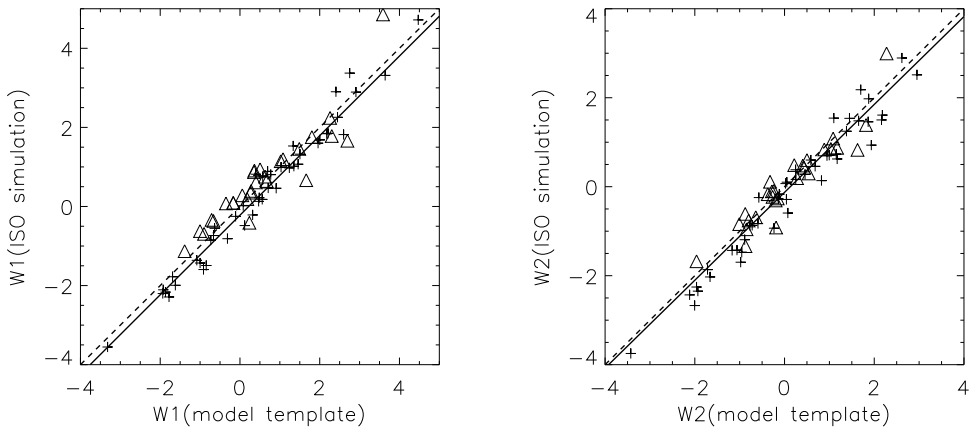}}
\caption{Comparison between the ISO simulation magnitudes and the simulated magnitudes from the model for AGB stars in the WISE W1 and W2 bands. Pluses are O-rich and triangles are C-rich AGB stars. Dashed lines indicate equality between the x- and y-axis values.}
\label{figure3}
\end{figure}

\section{Results}
\subsection{WISE calibration solution}
\begin{figure}
\resizebox{\hsize}{!}{\includegraphics{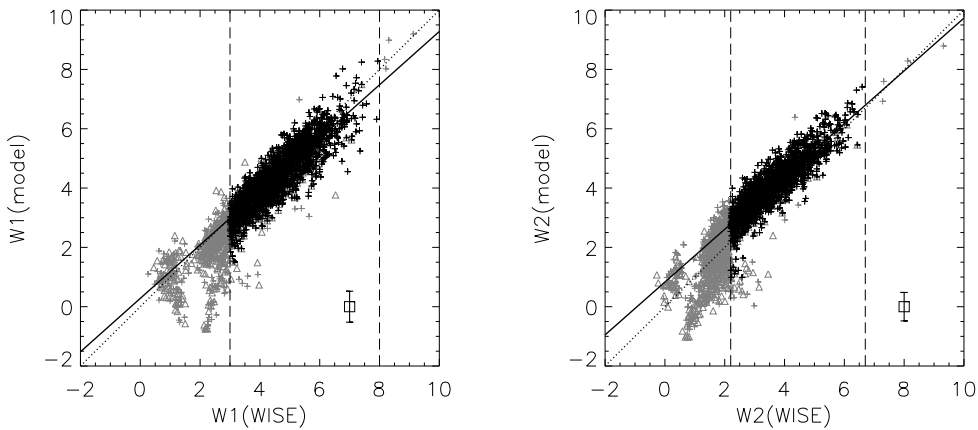}}
\caption{Comparison between W1 and W2 magnitudes from the best-fit
    models and WISE observations. Vertical dashed lines delineate the
  magnitude limits between which AGB stars are selected to derive
    the coefficients of calibration equations. Solid lines are the
  linear least-squares fits to the data points between lower
  limit and the WISE saturation limit. Dotted lines are the lines of the unit slope
  for reference. The error bars in the right corner indicate the median uncertainties in model magnitudes, including that caused by variability.}
\label{figure4}
\end{figure}

We show the comparison between the W1 and W2 model magnitudes and the WISE
observation magnitudes in Figure 4. A linear relationship seems to exist between
the model values and the observations for most sources at the faint
end, although these sources are all brighter than the nominal
saturation limits. Only a small fraction of the sources at the bright
end do not follow the relationships. The features at the bright end are
similar to those in Figure 1 when comparing ISO simulation magnitudes
with WISE. This also supports the consistency between ISO spectra and DUSTY models. The faint end vertical dashed lines indicate the WISE saturation limits (8 and 6.7 mag for W1 and W2). Vertical dashed lines at the bright end (3 and 2.2 mag for W1 and W2, respectively) indicate the criteria that are visually determined to exclude sources that disagree in the linear relationship. The bright useful limit is consistent with the WISE photometry limit mentioned in \textsection2.3. Twenty data points in the W1 and 32 data points in the W2 panels, which significantly deviate from the linear distribution, were excluded. The correlation coefficients of the data points selected by the criteria and WISE saturation limit are 0.898 and 0.910 in the W1 and W2 bands.

To acquire a more realistic relationship between model W1 and W2 and WISE W1 and W2, the uncertainties in the y-axis of the model synthetic magnitudes that are caused by variability of AGB stars need to be specified. We estimated the uncertainties caused by variability with the following three steps: first, we calculated a "typical" model SED with typical model parameters ($T_{\scriptsize{\textup{eff}}}=3000 K$, X=1, $T_{\scriptsize{\textup{in}}}=1000 K$ and $\tau_{\scriptsize{0.55}}=10$). Then, for each band of J, H, $\mathrm{K}_{\rm s}$, W3, and W4 of this "typical" model SED, we assumed a sinusoidal light curve with different amplitude and different phase. \citet{smith2002} suggested that the variation amplitude decreases with increasing wavelength. 
Since the variability amplitude of the semiregular variables (SRVs) is typically 0.4--0.8 mag and that of miras in the OGLE-III (Opitcal Graviational Lensing Experiment; \citealt{soszynski2013}) dataset are higher than 1 mag in the I band, we assumed the typical amplitude of variation as 0.7, 0.7, 0.7, 0.4, and 0.4 mag for J, H, $\mathrm{K}_{\rm s}$, W3, and W4 bands, respectively. We sampled each light curve at a random phase to generate a new SED. Finally, we simulated the new SED and compared the new W1 and W2 with the primary W1 and W2 of the "typical" model SED. We repeated the second and third steps to find the W1 and W2 model uncertainties, which are 0.522 and 0.480 mag.   

To determine the linear relation in a more robust way, we used median fitting. The data points were divided into several bins with 150 objects in each bin along the x-axis. For each bin we computed the median x- and y-values and used the median absolute deviation from the median (MADM) to estimate the spread in x- and y-values. Then we fit a line through these new binned data points. The results of the linear equation using median fitting are as follows:
\begin{eqnarray}
W1(model) & = & 0.28(\pm 0.83) \nonumber \\
& & +0.90(\pm 0.17)\times W1(WISE) \nonumber \\
W2(model) & = & 0.83(\pm 0.67)  \nonumber \\
& & +0.89(\pm 0.18)\times W2(WISE). 
\end{eqnarray}
The linear relations are plotted as solid lines in the Figure 4. The dotted lines of the unit slope are plotted for reference. The nonunity
slope of linear relationships indicate that the offsets
of the WISE observations are magnitude-dependent. Although several data
points near the bright criteria do not follow the linear relationship,
the fitted linear relations are valid for most data points. The
dispersion may be caused by the variability of AGB stars and
uncertainties in models. 
The relation between DUSTY models and
ISO observations as described in Equation 3 yields our final calibration formulas:
\begin{eqnarray}
W1(calibrated) & = & 0.10(\pm 0.91) \nonumber \\
&  & +0.98(\pm 0.19)\times W1(WISE) \nonumber \\
W2(calibrated) & = & 0.75(\pm 0.73)\nonumber \\
&  & +0.96(\pm 0.20)\times W2(WISE).  
\end{eqnarray}
The empirical calibration formulas are
valid only in the ranges of 3-8 mag for W1 and 2.2-6.7 mag for
W2. In our AGB star sample, the W1 and W2 magnitudes of 2390 and 2021 AGB stars are located in these ranges. 
The median of W(model)--W(PSF) is --0.024 and 0.557 mag for W1 and W2. It can also be deduced from Equation 5 that WISE PSF-fit photometry in general slightly underestimates the W1 flux and overestimates the W2 flux.

Several effects may contribute to the magnitude-dependent
offsets. First, AGB stars have extended profiles because of their dusty
mass-loss winds. The extended profile probably has an exaggerated
effect on the PSF magnitudes that are computed using the unsaturated
outer part of a saturated star image. Since brighter stars are
typically more extended, this effect might contribute to the nonunity slope. Second, variability of AGB stars will introduce uncertainties in model simulation and dispersion in Figure 4, which will reduce the slope. Finally, the flux-dependent shape of the PSF of the WISE detector in the saturated regime might be another contributor to the bright star flux biases, which are discussed in the section on photometric bias in the WISE Explanatory Supplement documentation of the All-Sky Data Release.

Since the uncertainties caused by variability of the y-axis are much larger than the uncertainties of observations in the x-axis, variability has a significant influence on the relation in Figure 4.
As we mentioned in \textsection2.3, the molecular absorbtion feature in the W1 and W2 bands may be an important reason for the DUSTY model deviation from ISO, some molecules such as CO also have features in the near-infrared band. It is hard to quantify the effect of the molecular feature on the model simulation. However, changes caused by the molecular feature ($\sim$ 0.1 mag in the W1 and W2 bands) to the infrared SED are much smaller than the change caused by variability ($\sim$ 0.5 mag in the W1 and W2 bands). Since we have taken variability uncertainties into consideration, adding uncertainties caused by molecular features will not make a noticeable difference to the current results.

\subsection{Separation between C-rich and O-rich AGB}

\begin{figure*}
\centering
\resizebox{\hsize}{!}{\includegraphics{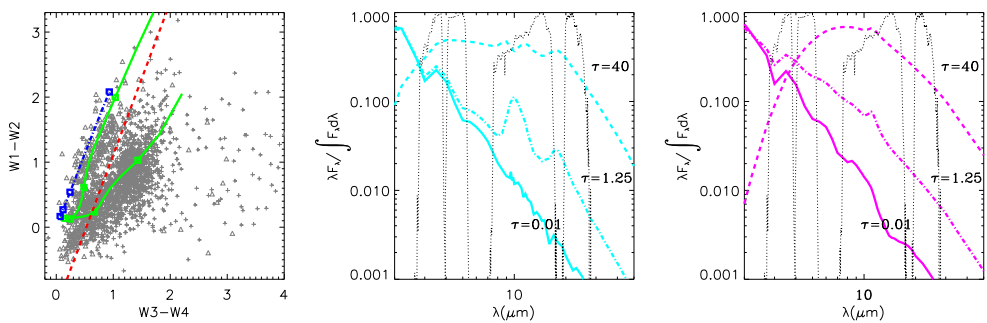}}
\caption{Left panel: WISE color-color diagram that can be used to distinguish O-rich and C-rich AGB stars. Both the W1--W2 color and W3--W4 color are computed from the calibrated photometry.
Pluses are O-rich stars, while triangles are C-rich stars. The dashed line is calculated to most effectively separate the two types of AGB stars. The green solid curves (left: O-rich, right: C-rich) are a series of models with different optical depth and typical values for other 3 parameter values. The dash-dotted line is a series of blackbodies for reference. The four blue squares from bottom to top are blackbodies with temperatures of 3500, 2500, 1500, and 500 K, respectively.The right two panels show the DUSTY model SEDs with optical depth $\tau_{0.55}$=0.01, 1.25, and 80 for O-rich AGB and C-rich AGB, respectively. The dotted curves are the WISE W1--W4 band relative spectral response functions. The simulated colors of these four models are designated as thick squares in the left panel.}
\label{figure5}
\end{figure*}

We plot the W1$-$W2 versus W3$-$W4 color-color diagram of our calibrated WISE sample in Figure 5. It can be seen that the two types of AGB stars are located in different regions in the figure and can be well distinguished in the WISE color-color diagram. A red solid line is plotted to effectively separate the two types of AGB stars. 
The straight lines successfully separate 87.1\% of the O-rich AGB stars and 85.7\% of the C-rich AGB stars.
We computed the separation line by varying the slope and intercept until the product of the two separation fractions were minimum. 
This line function is
\begin{equation}
W1-W2=2.35\times (W3-W4)-1.24 .
\end{equation}
It can be seen from Figure 5 that some data points lie far from the main distribution of AGB stars and show extremely high W3--W4 colors. They are the sample of objects whose SEDs cannot be successfully fit by our DUSTY model grids.

To probe the physical origin of this division, we calculated two
sequences of DUSTY models by varying the optical depth while
 keeping the other three parameters fixed
  ($T_{\scriptsize{\textup{eff}}}=3000$\,K, $X=1$ and
  $T_{\scriptsize{\textup{in}}}=1000$\,K) for both O-rich and C-rich
  AGB stars. The optical depth $\tau_{0.55}$ ranges
from 0.01 to 100. The resulting tracks are also plotted 
as green curves in the left panel of Figure 5. In general,
increasing the optical depth, which also indicates a decrease in the dust
temperature, causes the data point to move diagonally from left
bottom to upper right in the figure. O-rich models differ
significantly from C-rich models in that the W1$-$W2 color increases
much more slowly than W3$-$W4 does with increasing optical
depth. Therefore, the distribution of O-rich models moves almost
horizontally at the low W1$-$W2 end. The only difference between these
two sequences of models is the dust composition. We also plot the
  blackbody curve for temperatures from 500\,K to 3500\,K as the dash-dotted
  line in the figure. The distribution of carbon stars roughly
  follows the trend of the blackbody curve, but with slightly bluer
  W1$-$W2 or redder W3--W4 colors than the latter.

The middle and right panels of Figure 5 show the model SEDs with
optical depth equal to 0.01, 1.25, and 40 for O-rich and C-rich AGB stars
(middle panel for O-rich and right for C-rich) and WISE W1--W4 response
functions. The simulated WISE colors of these four models are
indicated as squares in the left panel of Figure 5. The
SED of the C-rich star model with an optical depth of 0.01 almost overlaps that of the
O-rich model with the same optical depth, since both of them are not
affected significantly by dust and are essentially the spectrum of the
central star. In contrast, SEDs of the models with an
optical depth of 1.25 and 40 are different between the two AGB
types, although the SEDs are normalized to the bolometric luminosities
of the stars. 
When $\tau=1.25$, the O-rich star model shows negligible
emission at the shorter wavelengths and thus is still dominated by the
central star emission, while the C-rich star model has begun to be
dominated by dust emission at these wavelengths. Conversely, O-rich AGB
stars show stronger dust emission at longer wavelengths than C-rich AGB
stars. This difference arises because O-rich dust species have much less efficient absorption
at shorter wavelengths but more efficient absorption at longer wavelengths than C-rich dust
species. This explains that the W1$-$W2 colors
of O-rich model AGB stars are not sensitive to the increase of optical
depth in the not very optically thick cases while their W3$-$W4 colors
redden faster than C-rich AGB stars. 
When the circumstellar envelope
becomes very optically thick (with $\tau=40$), the O-rich AGB star
model still has bluer W1$-$W2 and redder W3--W4 colors than the C-rich AGB star model. 
This is because
of the more efficient absorption at longer wavelengths and perhaps also because of the
self-absorption feature at 9.8 $\mu$m. 
The C-rich AGB stars are
located closer to the blackbody line because the C-rich dust species
have a relatively smooth opacity profile.

\subsection{Model parameters}

\begin{figure*}
\centering
\resizebox{\hsize}{!}{\includegraphics{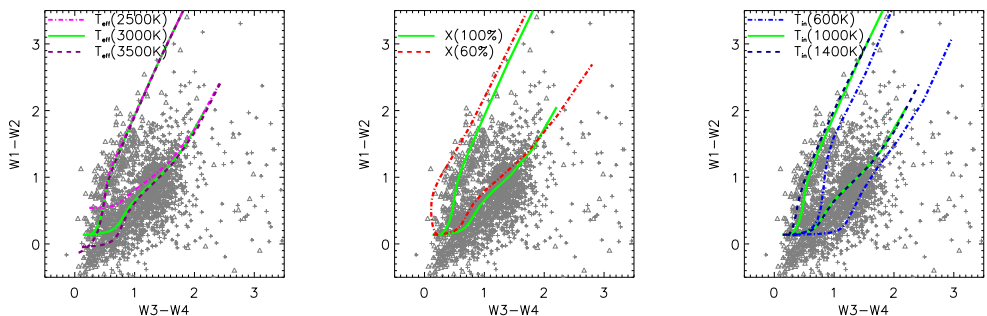}}
\caption{WISE color-color diagram overlaid with models with
  different parameters. Models with the three parameters
  $T_{\scriptsize{\textup{eff}}}$, X, and
  $T_{\scriptsize{\textup{in}}}$ are shown in different panels from
  left to right. Solid lines are the fiducial model
  series as shown in Figure 5. In each panel, the left set of tracks are models for C-rich stars and the right tracks plot O-rich models.}
\label{figure6}
\end{figure*}

From Figure 5, it can be seen that the trend of our WISE AGB
sample from left bottom to upper right in the color-color diagram is
predominantly caused by changes in optical depth. To analyze the effects of
 the other three model parameters
  ($T_{\scriptsize{\textup{eff}}}$, $T_{\scriptsize{\textup{in}}}$ and
  X) on the mid-infrared colors, we took the optical depth tracks in
  Figure 5 as standard model tracks and constructed new tracks for
  various values of the other three parameters. The three panels
  of Figure 6 compare the model tracks produced by varying one of the
  three other model parameters. 
Note that not all values of X might be physical. The maximum fractional abundance of SiC in the dust depends on the metallicity as well as on the fraction of Si that condenses into dust. The AmC-to-SiC ratio is also controlled by the dust condensation sequence; depending on whether C or SiC condenses first (see, e.g., Section 5 in \citet{leisenring2008} and references therein). These model curves are only used to show the range of variation in our models that is caused by changes in the model parameters. 
Solid lines are the standard model tracks as shown in Figure 5.

In the left panel of Figure 6, it can be seen that the change of
$T_{\scriptsize{\textup{eff}}}$ mainly affects the optically thin
objects. This is because the NIR SEDs of optically thin
AGB stars are dominated by the central star's radiation. In the middle
panel of Figure 6, the change of the dust composition ratio X mainly
affects the IR colors of C-rich AGB star models while its effects are quite
weak on O-rich AGB star models. This may be because the difference between the extinction curve of the two assumed dust components is steeper in C-rich AGB stars than that in O-rich AGB stars. In the right panel of Figure 6, models with higher $T_{\scriptsize{\textup{in}}}$ have fainter (i.e. bluer) W3--W4 and W1--W2 colors, which is consistent with the behavior of  the blackbody. 
The variety of these physical parameters in AGB
circumstellar envelopes contributes significantly to the data scatter in
the WISE color-color diagram.

\subsection{Implications of the number statistics} 

\begin{figure*}
\centering
\resizebox{\hsize}{!}{\includegraphics{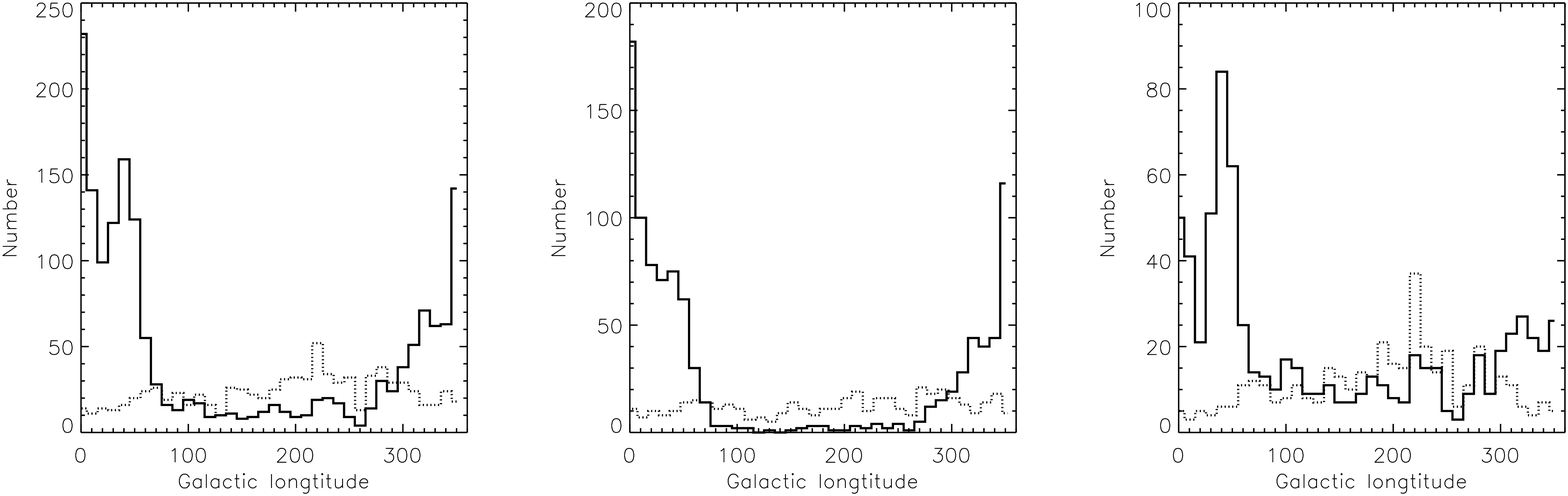}}
\caption{Longtitude distribution of our calibrated WISE AGB stars. The left panel shows all stars in the sample, while the middle panel shows objects with $\tau_{0.55}>10$ and the right panel shows stars with $\tau_{0.55}<10$. O-rich AGB stars are solid lines and C-rich stars are plotted as dashed lines.}
\label{figure7}
\end{figure*}

Because this is compilation of different works, it is hard to quantitatively
estimate the completeness of our WISE AGB sample. However, we can
determine the completeness from the spatial
distribution of the sources. Most of the WISE AGB stars are located at the
low Galactic latitude region, as expected. The longitude distributions
are shown in Figure 7. The left panel shows the whole sample, while
the right two panels show sources with optical depth $\tau_{0.55} > 10$
and $\tau_{0.55}<10$, respectively. It is obvious that more O-rich AGB
stars (solid) concentrate toward the direction of the Galactic center,
while C-rich AGB stars (dashed) are distributed more evenly along the
Galactic longitude. The different longitude distributions of
O-rich and C-rich AGB stars are consistent with previous works. For
example, \citet{Le2003} found that C-rich AGB stars are located
preferentially in the exterior of our Galaxy and O-rich stars tend to
reside in the interior of our Galaxy. The Galactic segregation of
these two types of AGB stars is usually interpreted as an effect of
the metallicity during the formation of C-rich AGB stars. Stellar
evolution models of AGB stars show that O-rich AGB stars can evolve
into C-rich AGB stars more rapidly {at} lower metallicity
\citep{Le2003}. Thus the metallicity
gradient in our Galaxy can cause the different longtitude
distributions of O-rich and C-rich AGB stars. The consistent spatial
distributions between the our sample and other works suggests that the
sample might represent a population of AGB stars in the solar neighborhood. This may be because the catalog is built from sources selected randomly in different works.

The middle panel of Figure 7 is the longitude distribution of AGB stars with $\tau_{0.55} > 10$ . Very few high optical depth O-rich AGB stars locate in the direction opposed to the Galactic center. This deficiency of high optical depth object is obvious from comparing the middle panel and the right panel of Figure 7. We can rule out the possibility that the differing Galactic interstellar extinction along different lines of sight could lead to this trend because no significant difference is present in the longitude distribution of low- and high optical depth C-rich AGB stars. The metallcity gradient in our Galaxy may again be a natural explanation because O-rich AGB stars in the exterior of our Galaxy have lower metallicty and will evolve into C-rich AGB stars at earlier evolution times. These O-rich AGB stars may have a relatively lower amount of dust in the circumstellar envelopes and lower optical depth.

\section{Summary}

We reported our findings on AGB star colors in the WISE
bands. We found that the WISE W1 and W2 magnitudes of AGB stars do not
agree with the spectroscopic measurements from ISO when we compared our
sample of ISO simulated W1--W4 magnitudes with WISE observations, which
we attribute to the residual bias in the PSF-fit photometry of
  saturated objects in these two bands. The WISE saturated W3 and W4 magnitudes are directly calibrated based on ISO synthetic photometry. To calibrate
  the WISE W1 and W2 bands, we resorted to ISO spectra of a
  subsample of our AGB stars and proved that the radiation transfer
  code DUSTY can be used to reproduce unbiased W1 and W2 magnitudes
  of bright stars. Using DUSTY, we successfully developed a calibration method
for the observed WISE W1 and W2 bands to lift the residual bias of the
PSF-fit photometry. The calibration procedure revealed that WISE may in
general have underestimated W1 flux and overestimated
W2 flux, and these deviations seem to be magnitude-dependent. 

Combining the model-calibrated W1 and W2 data with the directly calibrated WISE
  W3 and W4 magnitudes, we analyzed the W1$-$W2
vs W3$-$W4 color-color diagram and found that the two main types of AGB
stars, O-rich AGB and C-rich AGB, can be effectively distinguished by their
WISE colors. The division is mainly caused by the different extinction efficiencies between silicate-type and carbonaceous
grains. 
The spatial distribution of the AGB sample is consistent with
previous work. We also found that O-rich AGB stars with an opaque
circumstellar envelope are much rarer toward the anti-Galactic Center
direction than C-rich AGB stars, which we attribute to the
metallicity gradient of our Galaxy.

\begin{acknowledgements}
 
We thank the referee for thoughtful comments and insightful suggestions, which significantly improved the quality of this paper. This publication makes use of data products from the Wide-field Infrared 
Survey Explorer, which is a joint project of the University of California, 
Los Angeles, and the Jet Propulsion Laboratory/California Institute of 
Technology, funded by the National Aeronautics and Space Administration and the Two Micron All Sky Survey, which is a joint project of the University of Massachusetts and the Infrared Processing and Analysis Center/California Institute of Technology, funded by the National Aeronautics and Space Administration and the National Science Foundation. We are grateful to Martin Cohen, who provided the Vega spectrum to check our magnitude simulation process, and to Edward Wright, who gave us very useful advice about the origin of the magnitude-dependent offset of WISE photometry in the saturated regime. This work is supported by China Ministry of Science and Technology under State 
Key Development Program for Basic Research (2012CB821800), the National Natural Science Foundation of China (NSFC, Nos. 11173056, 11225315 and 11320101002), Chinese Universities Scientific Fund (CUSF) and Specialized Research Fund for the Doctoral Program of Higher Education (SRFDP, No. 20123402110037).

\end{acknowledgements}

\end{document}